\documentclass[useAMS,usenatbib,usegraphicx]{mn2e}
\usepackage{colortbl,color}
\usepackage{amstext}
\usepackage{multirow}

\title[Assessment of evolutionary status of eclipsing binaries]
{Assessment of evolutionary status of eclipsing binaries
using light-curve parameters and spectral classification}
\author[E. A. Avvakumova and O. Yu. Malkov]{E. A. Avvakumova$^{1}$
\thanks{E-mail: Ekaterina.Avvakumova@urfu.ru} and O. Yu.
Malkov$^{2,3}$\\
$^1$Kourovka astronomical observatory, Institute of Natural Science,
 B.N. Yeltsin Ural Federal University,\\
 19 Mira St., Yekaterinburg 620000, Russia\\
$^{2}$South African Astronomical Observatory, PO Box 9, Observatory 7935, South Africa\\
$^{3}$Institute of Astronomy of the Russian Acad. Sci., 48 Pyatnitskaya St., Moscow 119017, Russia}

\begin{document}

\date{Accepted 2014 August 1 /  Received 2014 August 1; in original form 2014 May 12}

\pagerange{\pageref{firstpage}--\pageref{lastpage}} \pubyear{2014}

\maketitle

\label{firstpage}

\begin{abstract}

We have developed a procedure for the classification of eclipsing binaries from
their light-curve parameters and spectral type. The procedure
was tested on more than 1000 systems with known classification, and its
efficiency was estimated for every evolutionary status we use.
The procedure was applied to about 4700 binaries with no classification, and
the vast majority of them was classified successfully.
Systems of relatively rare evolutionary classes were detected in that process,
as well as systems with unusual and/or contradictory parameters.
Also, for 50 previously unclassified cluster binaries evolutionary
classes were identified. These stars can serve as tracers for age
and distance estimation of their parent stellar systems.
The procedure proved itself as fast, flexible and effective enough to
be applied to large ground based and space born surveys, containing
tens of thousands of eclipsing binaries.

\end{abstract}


\section{Introduction}
Binary stars are numerous (from 50 to even 90 per cent in the local group). 
Doubled lined eclipsing binaries provide the only method by which fundamental stellar parameters
(such as mass, radius, luminosity, etc.) can be independently estimated
without having to resolve spatially the binary or rely on astrophysical
assumptions. Unfortunately, only a small fraction of all binaries eclipse,
and spectroscopy, with sufficient resolution, can only be performed for
bright stars. The intersection of these two groups leaves only several
hundred stars, an amount that is not growing significantly.

Meanwhile recent major advances in CCD detectors and the implementation
of image-difference analysis techniques enables simultaneous photometric
measurements of tens of thousands of stars in a single exposure, leading to
a dramatic growth in the number of stars with high-quality, multi-epoch,
photometric data. There are many millions of light curves available from
a variety of surveys, such as the ground based
ASAS \citep{Pojmanski2002}, MACHO \citep{Alcock1998}, 
OGLE \citep{Rucinski2001a}, EROS \citep{Grison1995},
TrES \citep{Alonso2004},
HAT \citep{Bakos2004} and the space born 
Kepler \citep{Matijevic2012}
and CoRoT \citep{Loeillet2008} projects.  Thus eclipsing binaries represent 
the most numerous type type of binaries with known orbital period, 
because it can be easily determined from the not very long photometric observational sets.
  However, the number of fully 
characterized eclipsing binaries has not grown
significantly, as there has not been a corresponding growth in the quantity
of spectroscopic data. Therefore it would be advisable to develop a procedure
for estimation of the fundamental parameters values for eclipsing variables with unknown
spectroscopic elements. Obviously, an assessment of eclipsing binary evolutionary status
should be performed prior to the start of the fundamental parameters
estimation, as the set of rules for parametrization varies from one evolutionary
status to another.

A procedure for determination of the evolutionary class from the rest of the
observational data was first proposed by \citet{Svechnikov1980}.
The procedure is based on
a restricted number of systems with known classes contained in
the old \citet{Svechnikov1969eng} catalogue which, as our analysis has shown
\citep{Malkov2006}, is not accurate enough.
Useful ideas for classification of eclipsing binaries can also be
found in a statistical study made by \citet{Giuricin1983}, however,
they mostly dealt with only three classes of systems (detached,
semi-detached and contact).

In this paper we present a novel procedure, which utilizes
the most comprehensive set of rules for the classification of
eclipsing binaries, while requiring only light curve parameters
and an estimate of the binary's spectral type or color index.
This procedure can be used to quickly characterize large numbers
of eclipsing binaries (which can be advisable e.g., for statistical investigations), 
and allows the user to categorize them, even if the set of the mentioned
parameters is incomplete.
The procedure was tested with the Catalogue of
eclipsing variables \citep[CEV,][]{Avvakumova2013}, which is the world's
principal database of eclipsing binary systems with available classification.

The scheme of classification is presented in Sect.~\ref{sec:class}. A 
testing and application of the procedure is described in Sect.~\ref{sec:test}. 
Discussion of systems with ambiguous or
contradictory classifications, as well as systems belonging to extreme
and unusual stages of evolution, can also be found in the section above.
In Sect.~\ref{sec:conclusion} we draw our conclusions. Appendix~\ref{app:examples} contains discussion of
selected binaries. In Appendix~\ref{app:application} we give an example of 
application of our classification procedure, while cluster binaries are listed in Appendix~\ref{app:cluster}.

\section{Classification scheme}\label{sec:class}

The main goal of our work is to develop a fast and
effective procedure for determination of the evolutionary status of eclipsing binaries.
Since 2004 \citep{Malkov2004} we have collected information on
light-curve parameters and other observational parameters of these variables, on one hand, and
recent published information about evolutionary status of eclipsing binaries, on the other hand. The second
version of the
Catalogue of Eclipsing 
Variables (CEV)\footnote{online live-version can be downloaded 
from http://www.inasan.ru/$\sim$malkov/CEV/}~\citet{Avvakumova2013}
contains about 7200 eclipsing binaries, and the
evolutionary status is available for about 1300 of them. 
The collected data allows us to make a preliminary statistical
analysis and find relations between the different parameters
for various evolutionary classes of eclipsing binaries. Such an analysis is
presented in this section.

Detailed description of the evolutionary classes used in the
current study can be found in~\citet{Avvakumova2013},
while meaning of the light-curve parameter designations, thought
generally accepted, is given in~\citet{Malkov2007}. 
The following data from CEV we used in the analysis:
\begin{itemize}
\item depth of  primary minima A$_1$, mag;
\item depth of secondary minima A$_2$, mag;
\item depth difference $\Delta$A=A$_1$-A$_2$, mag;
\item morphological type of the light curve (EA, EB, EW; as in
the GCVS);
\item period of the eclipsing variable star, P, days;
\item spectral type of the primary star, Sp$_1$;
\item luminosity class of the primary star;
\item spectral type of the secondary star, Sp$_2$;
\item luminosity class of the secondary star;
\item the components spectral type difference\\ $\Delta$Sp=Sp$_1$-Sp$_2$
\end{itemize}
Unlike to \citet{Malkov2007} we didn't use in our analysis
the information about variability of the period, data on duration 
of the eclipses and phase of secondary minimum. All these parameters are included in CEV, 
when available from literature.
However the number of such systems is relatively small, and additional
observations are required to enlarge that number. So we did not
include these parameters in the current version of our procedure.

\begin{figure}
\includegraphics[scale=0.5]{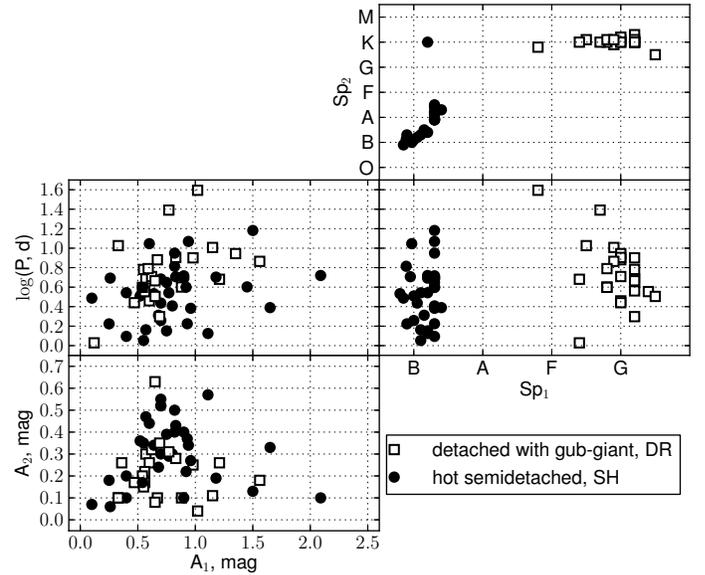} 
\caption{Location of hot semidetached binaries 
(SH, filled circles) and detached binaries with subgiants (DR, empty squares) 
in the A$_1$ (depth of primary minimum) -- A$_2$ (depth of secondary minimum)
-- P (period) -- Sp$_1$ (primary spectral type) -- Sp$_2$ (secondary spectral type)
planes.}
\label{SHvsDR}
\end{figure}

An example of the analysis is shown in
Fig.~\ref{SHvsDR}.
A distribution of two different evolutionary classes of binaries in the
A$_1$ (depth of primary minimum) -- A$_2$ (depth of secondary minimum)
is presented in Fig.~\ref{SHvsDR} (left bottom panel).
Hot semidetached binary class (SH, filled circles in Fig.~\ref{SHvsDR})
was introduced by \citet{Popper1980} in his review
to designate binary with the spectra of both components earlier
than classical algols spectra. About 30 such systems
are known. Empty squares in Fig.~\ref{SHvsDR} indicate detached 
subgiant systems (DR). All of these binaries are 
chromospherical active RS~CVn systems with the spectrum of the
primary of F-G~IV-V and with a strong H and K emission in the spectrum outside the
eclipse \citep{Hall1976}. Stellar activity is caused by the magnetic 
field on a star which is produced by the star's rapid rotation. 
The active component rotates faster then usual because of a spin up by
its close companion. According to \citet{Hall1981} the activity 
phenomena seen in the well detached RS~CVn binaries is fundamentally
different from those seen in the semidetached post-MS binaries, 
although a few semidetached RS CVn binaries are known  
\citep[see, e.g.,][]{Montesinos1988}. There are
about 20 such systems in the catalogue.

As can be seen in Fig.~\ref{SHvsDR} the depth of the primary minima A$_1$ of both SH and DR 
binaries is usually not larger than 1.5~mag. The value of the 
depth of secondary minima A$_2$ is generally not larger than 0.4~mag for DR 
systems, while A$_2$ of hot semidetached systems is not larger than 0.6~mag. 
Four exceptions are DR systems RW~UMa (A$_1=1.56$ mag), 
TY~Pyx (A$_2=0.63$ mag), and SH systems TT~Lyr (A$_1=2.09$ mag) and 
Z~Vul (A$_1=1.65$ mag). We have studied the literature available on these binaries
and their nature is discussed in the Appendix~\ref{app:examples}.

The distribution of the DR and SH systems in the parameter space 
$\log$P -- A$_1$ is shown on the top left panel of Fig.~\ref{SHvsDR}. 
We have found only three DR binaries with orbital periods being larger 
than $10$ days. All of them belong to the long period RS~CVn group. 
One DR binary (ES~Cnc) has unusual observational parameters namely 
short period and small A$_1$ value. We placed short description of this system 
to the Appendix \ref{app:examples}.

The left panels of Fig.~\ref{SHvsDR} 
clearly demonstrate that there is no difference between the two evolutionary
classes in the sense of values of depth of minima and the orbital period. 
Thus it is necessary to use additional observational data (such as 
information about chromospherical activity of RS~CVn systems, 
photometric distortion waves or/and variability of orbital period or 
information about the orbit eccentricity), if available, in order
to attribute a system to one or the other class.

Spectral type, which is known from observations or can be
estimated from the color indexes, can also serve as an additional parameter.
Secondary spectral type, when unknown, can be drawn from the
components' effective temperature ratio, which can be estimated from the
observed depths of minima $A_1, A_2$, if limb darkening is neglected: 
\begin{equation}
\frac{T_2}{T_1} = \sqrt[4]{\frac{J_2}{J_1}} = \sqrt[4]{\frac{1+0.4 A_2}{1+0.4 A_1}}.
\label{dmdm}
\end{equation}

Here $J_i$ is surface brightness of $i$-th component (see \citet{Brancewicz1980}
 and \citet{Malkov2012} for details).

Distributions of primary and secondary spectral types for the hotter and cooler 
components of DR and SH systems are shown on the right panels of Fig.~\ref{SHvsDR}.
It can be seen that even roughly estimated (e.g., from color indices) spectral type
allows us to separate detached subgiant systems from the hot semidetached systems.

We have performed such an analysis for every evolutionary class, and the
results are given in Table~\ref{parameters}. Evolutionary class
(with, in brackets, the number of such class systems in CEV) is followed by the
corresponding limits of observational parameters. These observational 
limits set the classification rules for assessment of the evolutionary classes. 
We list all the rules but the estimation of the evolutionary class is also 
possible when parameter set is incomplete. The limiting interval for both luminosity classes 
and morphological type of the light curve are additional parameters while the others are necessary.
We present the example of application of our classification method to one binary DP~CMa in the Appendix~\ref{app:application}.

\begin{table*}
\caption{The limits for observational parameters, used for the classification, for systems of different evolutionary class}
\label{parameters}
\begin{tabular}{@{}l p{20mm} c c c c c c l}\hline
Class$^1$&Description$^2$& A$_1^{~~3}$& (A$_2/\Delta$A)$^{3,4}$ & P& Sp$_1$  & Sp$_2$ & $\Delta$Sp$^5$ & MT$^6$\\
&&[mag]& [mag]&[days]&&&&\\\hline
\multicolumn{9}{l}{Detached binaries}\\[2mm]
DM (190)& MS stars & $1.10$ & $0.81$  & $[0.4;36]$ & O5--M4.5;IV-VI & O5--M4.5;IV-VI & up to 1.5 & EA, EB, E\\
DR (25)& with sub-giants  & $1.56$ & $0.35$  & $[1.9;26]$ & A8--G6;IV-V & G8--K3;IV-V & from 0.6 to 1.6 &  EA\\
DGE (8)& with OB giant, supergiant or WR star & $0.65$ & $0.34$  & $[1.6;35]$ & WR3--B2;I,III & O4--B3;III-V & up to 1.7 & EA,EB,E\\
DGL (16)& with late type giant or supergiant & $2.32$ & $0.20$  & $[69;7465]$ & B0--F7;I-V & G3--M2;I-III & from 1 to 4.5 & EA,EB,E\\
DW (14)& with WD & $6.00$ & $0.20$  & $[0.09;10]$ & WR8--B0;VI,wd & G8--M5;V,VI & from 4.8 to 6.5 & EA,EB,E\\
D2S (5)&symbiotic  & $6.22$ &  --$^{~~7}$ & $[603;6310]$ & WD,OB;V,wd & G5--M6;III & from 4.5 to 7.3 & EA,E\\[2mm]
\multicolumn{9}{l}{Semidetached binaries}\\[2mm]
SA (376)& classical Algols& $3.70$ & $0.60$  & $[2.1;45]$ & B4--G0;I-V & A2--M7;II-V & up to 3.8 & EA,EB\\
SC (5)& both late type stars & $1.36$ & $0.55$  & $[2.9;22]$ & G8--K4;III-V & K1--K5;III-V & from 0.1 to 0.5 & EA,EB\\
SH (34)& both early type stars  & $1.65$ & $0.57$  & $[1.1;16]$ & O8--B4;I,III-V & O9--A5;I-V & up to 1.2 & EA,EB,E\\
S2C (33)& cataclysmic & $6.00$ & $0.20$  & $[0.05;0.33]$ & WR5--B0;V,wd & G5--M9;V & from 4.5 to 6.9 & EA,EB,E\\[2mm]
\multicolumn{9}{l}{Contact binaries}\\[2mm]
CB (103)& near contact & $1.22$ & $0.81$  & $[0.2;1.5]$ & B8--K0;III-V & A0--M0;IV-V & up to 2.8 & EA,EB,EW\\
CBF (11)& F-subclass of CB& $1.00$ & $0.30$  & $[0.5;0.8]$ & A2--F4;V & G0--K3;IV-V & from 1 to 2.5 & EA,EB\\
CBV (13)&V-subclass of CB  & $0.91$ & $0.38$  & $[0.39;1.0]$ & A0--F8;V & F3--K5;V & from 0.8 to 2.6 & EA,EB\\
CE (19) & early type & $0.97$ & $0.28$  & $[0.49;1.9]$ & O7--B8;IV-V & O8--B8.5;IV-V & up to 0.5 & EB,EW,E\\
CWA (115)& late type, A-subclass & $0.81$ & $0.15$  & $[0.26;1.2]$ & A0--G8;III-V & A7--K0;V & up to 0.4 & EB,EW\\
CWW (123)& late type, W-subclass & $1.00$ & $0.22$  & $[0.22;0.78]$ & A7--K5;V & F8--K5.5;V & up to 0.5 & EB,EW\\
CG (4)& with early type giants or supergiants   & $0.69$ & $0.12$  & $[3.9;6.6]$ & O7--B0;I-III & WR9--B1;I-III & up to 0.3 & EB\\\hline
\end{tabular}\par
\medskip
$^1$the evolutionary status and the number of such systems in CEV;
$^2$for detailed description see \citet{Avvakumova2013};
$^3$maximum value;
$^4$A$_2$ value is given for DR, DG*, DW,
S* and CB*, 
and $\Delta$A value is given for DM, CE, CW* and CG (see text for details);
$^5$the components spectral type difference $\Delta$Sp=Sp$_1$-Sp$_2$
is given in units of a spectral class;
$^6$morphological type of the light curve;
$^7$data on secondary minimum are given in CEV for only one D2S system.
\end{table*}

CEV is photometrically heterogeneous, however, no magnitudes reduction
was made in the current study. Photometry for 97 per cent of 
the CEV eclipsing binaries is
given in one of the following four systems: $p$ (photographic),
$V$ (visual, photovisual, or Johnson’s $V$), $Hp$ (Hipparcos) and $B$
(Johnson’s $B$). In a first approximation we consider $p$ and $B$ as
being equivalent and the Hipparcos magnitude as not differing
much from {\it V} \citep{Bessell2000}. According to our estimations 
\citep{Malkov2007} $A_B /A_V = 1.07 \pm 0.01$ which 
leads to about 10 per cent inaccuracy in A$_i$ values. 

We believe that
the real interval limits should not differ significantly
from those given in Table~\ref{parameters}, as
they have astrophysical meaning.

An example for a detached main sequence (DM) binaries, where the primary
is larger, hotter and more massive than the secondary, was discussed
in \citet{Malkov2007}, where the following relation between
A$_1$ and A$_2$ upper limits was found:
\begin{equation}
A_1=-2.5\log\left(1-\frac{t^{2/\alpha}}{1+t}\right)+\sigma A_1,
\end{equation}
where $\alpha=5.5$ for MS stars from late O to 
early M, $t=10^{0.4A_2}-1$ and $\sigma A_1$ is an observational error
estimated to be about $0.3^m$.

Not all parameters are equally useful for the assessment of the
evolutionary status of the eclipsing binaries.
CW, CE, CG and majority of observable DM systems comprise similar components,
so the value of the depths difference $\Delta$A should not exceed some limit, and,
consequently, $\Delta$A value can serve as a good indicator of the evolutionary class.
Contrary, for DR, DG, DW, S, and CB systems, mostly comprising two
quite different components, we indicate a maximum value for the depth of
secondary minimum A$_2$.

\section{Procedure testing and application}\label{sec:test}

A large number of the newly discovered eclipsing variables have
an incomplete set of observational parameters. We have studied
the efficiency of our procedure and will discuss the main results
in the section below. 

\subsection{Membership probability}\label{sec:MP}

Our procedure should be effective and stable with respect to
the absence of some observational parameter values. In particular,
a lack of parameters leads to a condition when
a system resides in an area of the parameter space, covered
by two or more evolutionary classes. 
One example is described in Appendix~\ref{app:application}.
Another case is illustrated in two left panels of Fig.~\ref{SHvsDR} where both DR and SH classes 
can be assigned to binaries without known spectral types.

To solve this problem we calculate a membership probability 
(hereafter MP) for each class that can be assigned to 
the binary based on data of Table~\ref{parameters} by the classification procedure. The probability that
a given system belongs to a class $t$ ($MP_{t}$) is the 
ratio of binaries with available $t$-classification ($N_t$)
to the total number of binaries $\sum_i N_i$ with the available classification
in the $3\sigma$ radius around the examined system in the parameter space $S$:
\begin{equation}
MP_{t}=\frac{N_{t}}{\sum_i N_i},\quad i \in S_{3\sigma} \label{eq1}
\end{equation} 
We estimate $\sigma$ to
\begin{itemize}
\item $0.1$ mag for depth of minima. It is a typical photometric error for photographic
photometry, and at least half of magnitudes presented in CEV are photographic ones.
Other (mostly photoelectric) catalogued photometric data have a better accuracy;
\item about 25 per cent of period value itself.  These $\sigma$ leads to interval 
[0.25P;1.75P]. Although individual periods can in some cases 
be determined with very high precision, the choice of our $\sigma$ is
driven by the large range of periods in our training set data. Our testing has shown that adopting such a
large $\sigma$ does not degrade the performance of our results;
\item five spectral subclasses, which is an approximate accuracy of spectral type, estimated
from stellar color-index.
\end{itemize}

\begin{figure}
\includegraphics[scale=0.3]{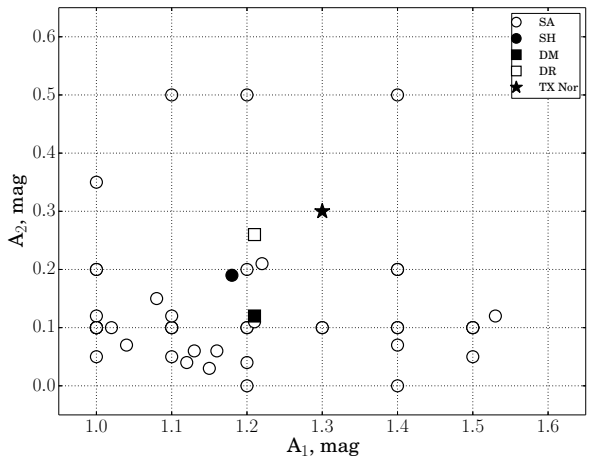}
\caption{On the calculation of membership probability.
The distribution of systems in the 2D-box parameter space.
Binaries with available classification are 
semidetached algols (SA, empty circles), 
hot semidetached (SH, filled circle),
detached main sequence (DM, filled square),
detached with subgiants (DR, empty square).
TX~Nor, the system of unknown evolutionary status, is indicated by the filled star.
See text for details.}\label{probability}
\end{figure}

This approach is illustrated in Fig.~\ref{probability}, which
represents a 2D-box (A$_1$--A$_2$) of the multi-dimensional
 space, where number of dimensions is
the number of observational parameters used for the classification.
The filled star is TX~Nor, the system of unknown evolutionary status, while
binaries with available classification are represented by
empty circles (semidetached algols, SA), filled circle (hot semidetached systems, SH), 
filled square (detached MS systems, DM)
and empty square (detached system with subgiants, DR).

Thus the membership probability for TX~Nor to be an
algol-like system equals to:
\begin{equation}
MP_{SA}=\frac{N_{SA}}{N_{SA}+N_{SH}+N_{DM}+N_{DR}}=\frac{32}{35}=0.91,
\end{equation}
$MP_{SH}$, $MP_{DM}$ and $MP_{DR}$ can be calculated similarly.
We consider the binary to be categorized successfully 
if the membership probability for one of the evolutionary classes
exceeds 0.5. 

The described procedure is simple, quick and can be implemented
in an automated program analysing published catalogues/lists of
eclipsing binaries.
However we should draw user's attention to the following features.

The majority of the catalogued systems with available classification 
(i.e. out training set)
have an incomplete parameter set. For example, spectral classification of 
both components is available only for 28 per cent of categorized CEV binaries. 
As a result, the smaller the parameter set that is available, the larger the number of
systems that are located within the given parameter space, and vice versa. If there are no systems in 
the $3\sigma$ vicinity the MP value can't be calculated 
because $\sum_i N_i$ in equation (\ref{eq1}) equals to zero. In such 
cases we increase the size of the parameter space by 
one or more sigmas while these number becomes greater than 1. 
However, the minimum size of $\pm3\sigma$ was sufficient to calculate MP for about 90 per cent of 
the investigated systems, and the maximum size of $\pm6\sigma$ was applied
to only two binaries.

The calculated MP values depends on available parameter set 
(i.e. the number of dimensions of the examined parameter space) 
thus we test our procedure on the systems with different parameter sets separately. 
This means that if a catalogued system has less or more parameters 
than examined unclassified binary it isn't included to the parameter space.

Another feature of the procedure is that MP value directly depends
on the number of different evolutionary class representatives in the vicinity
of an examined system.
For example, 
an area in the parameter space, occupied by SH systems,
is populated also by DM systems (see Table~\ref{parameters}).
However, the former evolutionary
class is much poorer represented among observed systems due to the
relatively small number of (high mass) objects and a rather rapid stage
of stellar evolution. Consequently, DM systems are more numerous
in the parameter space, and an examined system will be 
categorized as a DM system with higher probability.
We believe it is a correct solution, as
the examined system will more likely belong to an evolutionary class
of frequent occurrence.

Nevertheless we have published at CEV all 
the predicted classes to each of the considered binary 
not only the predicted class with the higher MP value.

\subsection{Efficiency of the procedure}

To estimate the efficiency of the procedure we have applied it to
CEV binaries with already available classification.
The results are given in the second column of Table~\ref{class}.
Data are presented separately for every parameter set, used for classification
(the first column).
The second column contains the
total number of CEV systems with a given parameter set followed by the
percentage of correctly classified binaries and, in brackets,
the percentage of unclassified binaries whose evolutionary class remained unknown
(i.e., we can't assign any of the classes to the binary based on its observational parameters).

Data for systems with no available classification
(see Sect.~\ref{sec:AP}) are presented in the
third column in a similar format, but numbers of successfully classified
(instead of correctly classified) systems are given here.
The binary has been considered to be successfully classified if
the membership probability for one of the possible classes was
larger than 0.5.

\begin{table}
\caption{The efficiency of the classification procedure for the CEV systems 
with different parameter sets}
\label{class}
\begin{tabular}{l m{2.5cm} m{2.5cm}}\hline
&Class is known& Class is unknown\\\cline{2-3}
Parameter set & Total number in CEV; correctly classified (unclassified)&
 Total number in CEV; successfully classified (unclassified)\\\hline
A$_1$, A$_2$, P, Sp$_1$, Sp$_2$ & 327; 81\% (1.5\%)& 59; 88\% (12\%)  \\
A$_1$, A$_2$, P, Sp$_1$         & 437; 65\% (4.6\%)& 868; 85\% (6\%) \\
A$_1$, A$_2$, Sp$_1$, Sp$_2$    & 0; & 0; \\
A$_1$, A$_2$, P                 & 278; 56\% (3.6\%)& 1359; 79\% (2\%) \\
A$_1$, A$_2$                    & 0; & 12; 75\%  (0\%)\\
A$_1$, P, Sp$_1$, Sp$_2$        & 75; 84\% (5\%)&  40; 73\% (25\%) \\
A$_1$, P, Sp$_1$                & 197; 73\% (1.5\%)&  395; 79\% (10\%)\\
A$_1$, P                        & 78; 42\% (0\%)& 1584; 84\% (2\%)  \\
A$_1$, Sp$_1$, Sp$_2$           & 1; 0\% (0\%)& 15; 80\% (13\%)\\
A$_1$, Sp$_1$                   & 12; 67\% (0\%)&371; 79\% (7\%) \\\hline
\end{tabular}
\end{table}

As can be seen from the second and third columns of Table~\ref{class} 
if spectra and period are known then efficiency of our procedure exceeds 80 per cent.

We have also estimated the efficiency of the procedure for each
evolutionary state. Results of the application of the same procedure to CEV systems with
available classification are given in the error matrix (Table~\ref{conmatrix}).
The first column contains CEV evolutionary classes,
each row of the table gives the result of the classification.
For example, the first row indicates that among all 190 CEV DM-systems
(see Table~\ref{parameters}),
174 were correctly categorized as DM, one was wrongly categorized as DR, etc.

So the matrix diagonal contains numbers of correctly categorized systems,
while the other cells of the matrix contain
numbers of wrongly categorized systems.
The last column contains false negative (type II) error values,
and the bottom row contains false positive (type I) error values.

\begin{table*}
\setlength{\tabcolsep}{2mm}
\caption{The confusion matrix of classification procedure}\label{conmatrix}
\begin{tabular}{c|lcccccccccccccccr}\hline
&\multicolumn{15}{c}{Results of classification}&\\
\multirow{17}{3 mm}{CEV\\class}&    &DM  &DR  &DGE &DGL&DW  &D2S&SA  &SC &SH &S2C &CB  &CE &CWA &CWW&CG &type II\\      
&            &                        &    &    &   &    &   &    &   &   &    &    &   &    &    &  & err.,\%\\ \hline 
&DM          &\cellcolor[gray]{0.8}174& 1  &    &   &    &   & 11 &   & 1 &    &  3 &   &    &    &  &7\\
&DR          & 8  &\cellcolor[gray]{0.8}10&    &   &    &   & 6  &   &   &    &    &   & 1  &    &  &60\\
&DGE         &    &    &\cellcolor[gray]{0.8}6&   &    &   &    &   & 2 &    &    &   &    &    &  &25\\
&DGL         &    &    &    &\cellcolor[gray]{0.8}13&    &   & 1  &   &   &    &    &   &    &    &  &19 \\
&DW          &    &    & 1  &   &\cellcolor[gray]{0.8}10&   &    &   &   & 1  &    &   &    &    &  &29\\
&D2S         &    &    &    &   &    &\cellcolor[gray]{0.8}5&    &   &   &    &    &   &    &    &  &0\\
&SA          & 19 &    &    &   &    &   &\cellcolor[gray]{0.8}330&   &   &    & 16 &   & 1  &    &  &12\\
&SC          & 2  &    &    &   &    &   &    &\cellcolor[gray]{0.8}3&   &    &    &   &    &    &  &40\\
&SH          & 15 &    & 1  &   &    &   & 1  &   &\cellcolor[gray]{0.8}16&    &    &   &    &    &  &53\\
&S2C         &    &    &    &   & 3  &   &    &   &   &\cellcolor[gray]{0.8}27& 1  &   &    &    &  &18\\
&CB          & 3  &    &    &   &    &   & 9  &   &   &    &\cellcolor[gray]{0.8}87&   & 4  & 4  &  &31\\
&CE          & 3  &    &    &   &    &   &    &   &   &    &    &\cellcolor[gray]{0.8}8& 1  &    &  &58\\
&CWA         & 2  &    &    &   &    &   &    &   &   &    & 4  &   &\cellcolor[gray]{0.8}80& 21 &  &30\\
&CWW         &    &    &    &   &    &   &    &   &   &    & 1  &   & 28 &\cellcolor[gray]{0.8}91&  &26\\
&CG          &    &    &    &   &    &   &    &   & 1 &    &    &   &    &    &\cellcolor[gray]{0.8}2&50\\\hline
type I  &   &23&9 &25&0  &23&0  &8 &0  &20 &4 &40&0  &30&21&0 &\\
err.,\%    &   &   &     &    &   &    &   &    &   &   &    &    &   &    &    &  &\\\hline
\end{tabular}
\end{table*}

Our analysis of the results, presented in Table~\ref{conmatrix}, shows
that the availability of different observational parameters can be crucial
for the classification of binaries of various evolutionary classes.
The following conclusions can be made.

In the case of detached systems with subgiants (DR), data about secondary 
spectra is required for correct classification, so only about 
40 per cent of these systems have been classified correctly. A reliable
classification of detached systems with OB giants (DGE) is only possible
when the luminosity class is known because all other observational 
parameters values are virtually the same for DGE and DM systems.
Detached
systems with white dwarfs (DW) are close to cataclysmic semidetached
binaries except the more longer periods. Therefore short-period DW
systems may be mis-classified as S2C or as detached systems with 
OB giant, if secondary spectra are unavailable. Our procedure is efficient 
for detached symbiotic systems (D2S) and detached systems with late type giants 
(DGL) because of the large luminosity difference between the components and their
long orbital periods.

Among semidetached systems of different classes the highest efficiency
is for semidetached algol-like binaries (SA)
and the lowest one is for hot semidetached systems (SH). The percentage of correct 
classification for cataclysmic semidetached systems is independent 
on the parameter set and is about 80 per cent.

Our procedure exhibits
the lowest efficiency for all classes of 
contact binaries because their observational parameters are close to 
parameters of detached MS systems and semidetached classical algol-like 
systems. Also, the procedure can not separate
CBF systems from CBV ones, however, it correctly identifies most of them
as near-contact CB binaries. Our procedure allows to 
separate the two subclasses of W~UMa systems since 70 per cent of CWA 
and 76 per cent of CWW binaries have been classified successfully.

\subsection{Application of the procedure}
\label{sec:AP}

After the testing, the procedure was applied to CEV systems
with no available classification. 
Before application we have checked the overlapping of training and 
prediction sets and found that density distribution of the parameters of both sets are the same.

The resulting statistics of application the procedure is given
in the third column of Table~\ref{class}.

We have detected a large number of candidates for interesting evolutionary 
classes, requiring further observations and studies. In particular,
we have indicated a number of candidates for detached systems:
74 of them are suspected to consist of a white dwarf and an OB companion;
36 of them are presumably MS systems with at least one OB massive component
and 30 others are presumably MS systems with a late-K or M star.
Three new candidates for cataclysmic systems (S2C) were also found.
 
Determination of the
basic stellar parameters of the components
of cluster or nearby galaxy binaries allows us
to measure the ages and distances of their parent stellar system,
and to test stellar evolution models
\citep[see][]{Graczyk2014,Rucinski2005}.
Based on our results and data from Simbad database 
we have compiled 
a list of cluster binaries with known evolutionary status
(see Table~\ref{clustermembers}). 
The list includes only previously unstudied binaries, with the
evolutionary class determined via our procedure.

We have checked all systems with unusual parameter values during the
application of our procedure. For most of them those
values are obsolete or unconfirmed. New observations of those binaries
are needed.
Another reason of unsuccessful
classification is a marginal (usually well-known) evolutionary status of a system.
SX Cas (``active algol'') can serve as a good example.
The third reason of unsuccessful
classification is the contradictory parameter values,
i.e., some observational parameters point to
one class, while others point to another.
One of such binary RT~Lac is described in 
Appendix~\ref{app:examples}. However for the majority of such systems
we failed to find in literature a reasonable explanation 
for contradictory parameters' values, and their
nature remains unclear.

We have compiled and published
lists of systems, belonging to these three categories, in the recent version of CEV.

In some cases an either too small or too large period value prevents
successful classification of the system. Period of an eclipsing binary with
negligible secondary minimum can erroneously be determined (and catalogued) to be
twice longer than the real one. Contrary, catalogued period of a binary
with an equal or similar minima can be twice shorter than the real value.
Our procedure can detect such cases. For example binary VW~Hya
has orbital period P$=2.69$ days and depth of the primary minimum A$_1=3.12$ mag 
in our catalogue. We have taken photometric data from \citet{Burki2005} 
and period  was given by \citet{Kreiner2004}. With these values of 
period and A$_1$ value of depth of secondary minimum A$_2$ equals to zero
and binary can be classified as classical algol-like (SA) system.
But \citet{Pojmanski2002} has given twice longer period. 
In this case A$_2$ equals to A$_1$ and our method can't classify 
binary as semi-detached algol-like system.

Such detection of half/double period confusion is only possible in cases where the wrong period
does not produce a predicted evolutionary class.

The analysis also shows that the procedure can indicate errors
in catalogued data.
In particular, we have detected and removed from CEV
(after confirmation from literature) about 30 non-eclipsing variables.
All these 30 objects weren't classified with our procedure.

Catalogue of eclipsing variables CEV and the results 
of the classification
are available in CDS VizieR service. Live version of the data can be
downloaded from http://www.inasan.ru/$\sim$malkov/CEV/

\subsection{Systems with uncertain or tentative classification}

CEV contains a number of eclipsing binaries
with an uncertain or tentative evolutionary class.
Examples are MU~Aqr (CB:) and TU~Boo (CWW and CWA, according
to different sources).
Most of such binaries were taken from lists of \citet{Shaw1994} 
and \citet{Pribulla2003}, and we have not found any other confirmation 
of the assigned evolutionary class(es).

We have applied our procedure to these binaries, and presented the results
in Table~\ref{doubleclass}. The second and third columns list
evolutionary class from CEV and one, determined with our procedure, respectively.
The fourth column contains the MP value, or a \verb|*| flag,
if the system was classified unambiguously.
In the fifth column we have used letter 'L' 
to refer to binaries without available light and radial curve analysis. 
Letter 'M' denotes binaries with contradictory classification.
Reference to the source of CEV evolutionary class is given in the last column.

\begin{table*}
\caption{The list of binaries with tentatively known evolutionary state and results of their classification}
\label{doubleclass}
\begin{tabular}{l l l c l l|l l l c l l}\hline
\multirow{3}{2 mm}{GCVS\\name}&\multirow{3}{2 mm}{CEV\\class}&\multirow{3}{8 mm}{Deter-mined class}&\multirow{3}{*}{MP, \%}&\multirow{3}{*}{Note}&\multirow{3}{*}{Ref}&\multirow{3}{2 mm}{GCVS\\name}&\multirow{3}{2 mm}{CEV\\class}&\multirow{3}{8 mm}{Deter-mined class}&\multirow{3}{*}{MP, \%}&\multirow{3}{*}{Note}&\multirow{3}{*}{Ref}\\
&&&&&&&&&&\\
&&&&&&&&&&\\\hline
CN And    &CWA,CB   & CB  &  * & M &(1),(2)&VY Lac    &SA,CB    & CB  & 45 & M &(13),(3)\\
EE Aqr    &CBV (:)  & CB  & 63 & M &(3),(4)&CN Lac    &CB,S     & CB  & 94 & L &(6),(14)\\
MU Aqr    &CB  (:)  & CWW & 56 & L &(5)&UV Leo    &CB,DM    & CB  & 80 & M &(14),(15)\\
RX Ari    &CB,S     & CB  & 53 & L &(6),(7)&RT LMi    &CWW,CWA  & CWA & 46 & M &(16),(17)\\
EP Aur    &CB  (:)  & CB  & 68 & L &(3)&V Lep     &CB  (:)  & CB  & 41 & L &(3)\\
TU Boo    &CWW,CWA  & CWW & 60 & M &(8),(9)&RR Lep    &SA,CB    & CB  & 52 & M &(18),(3)\\
DU Boo    &CWA (:)  & CB  & 42 & M &(10),(5)&KQ Lib    &CB  (:)  & CWA & 58 & L &(5)\\
RV CVn    &CW,CB    & CB  &  * & L &(11),(5)&V574 Lyr  &CB  (:)  & CWW & 60 & L &(5)\\
CW CMi    &CB  (:)  & CWW & 50 & L &(5)&FR Ori    &SA,CB    & SA  & 88 & M &(19),(3)\\
AL Cas    &CE  (:)  & CE  &  * & L &(5)&VZ Psc    &CB,CWA   & CWA & 92 & M &(20)\\
GS Cep    &CB  (:)  & SA  & 47 & L &(3)&EE Psc    &CB  (:)  & CWA & 50 & L &(5)\\
V628 Cyg  &CE  (:)  & CE  & 53 & L &(5)&VY Pup    &SA  (:)  & CB  & 66 & L &(3)\\
V680 Cyg  &CB  (:)  & DM  & 60 & L &(3)&V525 Sgr  &CBV (:)  & CB  & 55 & L &(3)\\
V1034 Cyg &SA  (:)  & CB  & 39 & L &(3)&RS Ser    &CB  (:)  & CWA & 46 & L &(5)\\
TZ Dra    &SA  (:)  & CB  & 42 & L &(3)&CQ Ser    &CB  (:)  & CB  & 61 & L &(3)\\
MT Her    &CB,S     & CB  & 83 & M &(3)&V Tri     &SA  (:)  & SA  & 57 & L &(3)\\
V1055 Her &CB  (:)  & CWW & 81 & L &(5)&AW Vir    &CWA,CWW  & CWW & 54 & M &(21),(22)\\
RS Ind    &CBF (:)  & CB  & 62 & M &(3)&KZ Vir    &CWA,CB   & DM  & 88 & L &(23),(5)\\
RY Ind    &SA,CB    & CB  & 57 & M &(12),(3)&CD Vul    &SA,CB    & CB  & 84 & L & (3),(24)\\\hline
\end{tabular}
\par
\medskip
(1) \citet{Jassur2006}; (2) \citet{vanHamme2001}; (3) \citet{Shaw1994};
(4) \citet{Wronka2010}; (5) \citet{Pribulla2003}; 
(6) \citet{Dryomova2005}; (7) \citet{Budding2004}; 
(8) \citet{Niarchos1996}; (9) \citet{Coughlin2008}; 
(10) \citet{Djurasevic2013}; (11) \citet{Schilt1927}; 
(12) \citet{Lapasset1982}; (13) \citet{Semeniuk1984}; 
(14) \citet{Giuricin1983b}; (15) \citet{Frederik1996}; 
(16) \citet{Niarchos1994}; (17) \citet{Rucinski2000}; 
(18) \citet{Vyas1989}; (19) \citet{Zakirov1996}; 
(20) \citet{Hrivnak1995}; (21)\citet{Lapasset1996}; 
(22) \citet{Niarchos1997}; (23) \citet{Rucinski2001}; 
(24) \citet{Brancewicz1980}  
\end{table*}	

As can be seen from Table~\ref{doubleclass}, three systems
were classified unambiguously, namely CN~And, RV~CVn and 
AL~Cas. For the last two systems there is no confirmation of our results 
in the literature because both systems have never been properly 
studied. Our evolutionary class may be helpful for such investigations. 
The near-contact evolutionary class for CN~And was confirmed by
light curves properties (e.g. asymmetry of maxima and unequal 
depth of minima) and by the solution of light curves which have been 
derived by \citet{vanHamme2001}. 

For 15 binaries in Table~\ref{doubleclass} tentative evolutionary class was 
confirmed by our procedure. 
For EE~Aqr, RS~Ind and V525~Sgr
near-contact evolutionary class (CB) was confirmed while
a subclass (CBV or CBF) remained unknown.
MP value for VY~Lac, RT~LMi and V~Lep
of the calculated evolutionary class
is smaller than 50 per cent, but our classification is correct.

VY~Lac, besides near-contact (CB) system, may also be
a semidetached algol-like system (MP = 34 per cent) or a detached
MS system (MP = 21 per cent).
The uncertainty in RT~LMi evolutionary class actually remains
as MP value for CWA class is only two percent larger than the one for CWW class.
V~Lep, besides CB system, may also be classified as a detached MS system
(MP = 38 per cent).

For seven binaries in Table~\ref{doubleclass} the determined
evolutionary class differs from the CEV (tentative) one.
All of these binaries have never been studied 
carefully, thus our classification can be considered as a proper one
until new observational data are obtained.
For example, KZ~Vir was classified as an A-type W~UMa contact 
system by \citet{Rucinski2001}, but they have noted that system may be 
a close detached binary. Later \citet{Pribulla2003} denoted the system 
as CB. So our DM class is probably correct, as
MP value for DM class is close to 100 per cent, and neither CWA nor CB class 
were assigned to this system by our method. 

The remaining seven systems were also classified but MP values 
were smaller than 50 per cent.
DU~Boo may be classified as a near-contact system with MP = 42 per cent
and as a detached MS system with MP = 41 per cent but not as 
a contact W~UMa binary.
CW~CMi was denoted as a near-contact by 
\citet{Pribulla2003}, but it is rather a contact W~UMa of W-subtype 
(with MP = 50 per cent) or A-subtype (with MP = 45 per cent).
The same situation applies to EE~Psc,
which is CWA or CWW system with MP value of 50 and 45 per cent, respectively, 
whereas MP value for the system to be a near-contact, as \citet{Pribulla2003} 
have supposed, is only 5 per cent.
We also have found that RS~Ser is CWA or CWW system
with corresponding MP values much larger than for near-contact configuration.
GS~Cep appears to be a semidetached SA system while 
MP value for near-contact class is smaller and equals to 23 per cent.
In contrast, V1034~Cyg is rather a near-contact binary with MP = 39 per cent vs. 
MP = 21 per cent for the semidetached class.
For TZ~Dra we have 
derived two possible classes, namely the near-contact (MP=42 per cent) and 
the semi-detached SA with an almost equal probability MP=41 per cent.

All the systems except DU~Boo were not previously studied, so our results can
be useful for future investigations.

\section{Conclusions}\label{sec:conclusion}

We constructed a procedure for the classification of eclipsing binaries,
based on light-curve parameters and spectral classification (or color-indices).
The procedure uses relations between different observational parameters
and allows us to attribute a binary to one of the 15 evolutionary classes
and estimate a probability of a correct classification.
We tested the procedure, using about 1000 binaries with available classification,
estimated its efficiency for different evolutionary classes
and applied it to 4700 systems with no classification, listed in
the Catalogue of eclipsing variables.
About 3800 systems were successfully classified. About 100 of them
happened to belong to some relatively rare evolutionary classes and
could be interesting for a further study.
Other 50 binaries, with newly determined evolutionary classes,
are members of stellar clusters and can
be used as additional tracers for age and distance estimation
of their parent stellar systems.

At the same time observational parameters of about 360 systems
are too unusual and/or contradictory to provide successful classification.
Published data for the most of such systems are obsolete or unconfirmed,
and new observations of these objects are needed.
Some other binaries are well-known to belong to a marginal evolutionary
status, while the nature of the rest 50 stars remains unknown.
About 40 catalogued systems with uncertain or tentative classification were
successfully classified with our procedure. Errors in catalogued data
can also be indicated: in particular, some 30 non-eclipsing variables
were found and, after confirmation from literature, removed from CEV.

The procedure is fast, effective and can be applied to eclipsing binaries
even if a set of observational parameters is incomplete.
It can be extremely useful
for the classification of a huge number of objects in large
ground based (MACHO, OGLE, etc.) and space born (Kepler, CoRoT, Gaia)
surveys.

Catalogue of eclipsing variables CEV and the results of the classification
are available in CDS VizieR service. Live version of the data can be
downloaded from http://www.inasan.ru/$\sim$malkov/CEV/

\section*{Acknowledgments}
We are grateful to Dmitry Kniazev and the reviewer,
Berry Holl whose constructive comments
greatly helped us to improve the paper.
This work has been partially supported
by Russian Foundation for Fundamental Research grants 12-07-00528 and 14-02-31056
and by the Presidium RAS program ``Leading Scientific Schools Support'' 3620.2014.2.
This research has made use of the SIMBAD database, 
operated at the Centre de
Donn\'ees astronomiques de Strasbourg,
and NASA's Astrophysics Data System Bibliographic Services.

\def\apj{ApJ}
\def\apjl{Astrophys.~J.,~Lett}
\def\an{Astron.~Nachr}
\def\aap{A\&A}
\def\mnras{MNRAS}
\def\pasp{PASP}
\def\aaps{Astron.~and Astrophys.,~Suppl.~Ser}
\def\apss{Astrophys.~Space.~Sci}
\def\ibvs{Inf.~Bull.~Variable~Stars}
\def\japa{J.~Astrophys.~Astron}
\def\na{New~Astron}
\def\aspproc{Proc.~ASP~Conf.~Ser.}
\def\aspcs{ASP~Conf.~Ser.}
\def\aj{AJ}
\def\actaa{Acta Astron}
\def\araa{Ann.~Rev.~Astron.~Astrophys}
\def\caosp{Contrib.~Astron.~Obs.~Skalnat{\'e}~Pleso}
\def\pasj{PASJ}
\def\memsai{Mem.~Soc.~Astron.~Ital}
\def\astl{Astron.~Letters}
\def\aipproc{Proc.~AIP~Conf.~Ser.}

\label{lastpage}

\clearpage
\appendix
\section{Discussion of selected binaries}\label{app:examples}

TU~Boo may be classified as a contact W~UMa of A-subtype because the primary minimum is a 
transit. The asymmetry of light curves was detected by several 
authors \citep{Niarchos1996,Coughlin2008}. \citet{Niarchos1996} found 
the light curve solution with WD program using spotted model for
contact configuration. They stressed that some physical 
characteristics of the binary (e.g. mass ratio) are typical for a W-subtype system. 
Moreover, their solution with spots shows that
the less massive and smaller secondary is hotter than the primary. 
According to \citet{Coughlin2008} TU~Boo is a marginal contact system 
with both components almost filling their critical lobes. They supposed 
the mass transfer from secondary to primary which is supported 
by an increased period.
Physical parameters (temperature ratio, radii ratio and masses),
derived by \citet{Coughlin2008}, point to
A-subtype, but the small percentage of overcontact (which leads to 
marginal contact only) together with q$\approx0.5$ appropriates 
to W-subtype. 

which has such small values of A$_1$ and period. 
According to \citet{Yakut2009} eclipses are partial. 
Additionally system is a hierarchical triple in which 
all three stars are blue stragglers.
 
RT~LMi was classified by \citet{Niarchos1994} as W-subtype of W~UMa 
systems based on the solution of observed light curves while 
\citet{Rucinski2000} assigned it A-subtype based on derived 
radial curves. Recently \citet{Qian2008} have shown that the primary 
minimum changed from occultation to transit and concluded that for RT~LMi 
a subtype based on Binnendijk's classification could not be uniquely assigned.
 
RT~Lac is one of the promising examples of the contradictory 
classification. The observed value of the secondary minimum depth
and the primary spectral type are not typical compared to other SA systems.
We have tried to classify it as a detached 
RS~CVn system, but A$_2$ value is not typical for DR class too.  
Moreover the evolutionary state of binary is not known exactly because
RT~Lac is among the most peculiar stars of RS~CVn type systems.
While most of RS~CVn binaries have equal-mass components, the
components of RT~Lac do not. \citet{Ibanoglu2001} reported 
that the brightness of the system at three phases, i.e., mid-primary and
quadratures, shows quasi-periodic changes which are caused by a 
chromospheric activity of a more
massive, smaller and hotter component. Moreover \citet{Ibanoglu1997s} 
showed that the less massive, 
larger star fills its critical lobe. Therefore, a gas stream from the
larger, less massive star to the more massive one will be expected. 
The binary may also belong to cool semidetached systems (SC), but its 
period is smaller than for other SC systems. 

TT~Lyr was classified as a hot semidetached system because of the spectral 
type of the primary, but the secondary (cooler) spectral type
is K0, according to \citet{Liao2010}. There is 
no comprehensive analysis of the photometric and spectroscopic data for 
TT~Lyr in the literature.
 
TY~Pyx is a unique active binary of RS~CVn type because as
\citet{Andersen1975} have shown, it consists of two almost identical
components. \citet{Rao1981} have supposed that both components are
on the pre-MS contraction phase.

RW~UMa is a detached system with an evolved subgiant component according
to \citet{Popper1977}. The value of A$_1$ is confirmed by NSVS data
\citep{Wozniak2004}. Thus we have used the value $1.56^m$ as the 
upper limit for the depth of the primary minimum for DR binaries.

Z~Vul is a hot semidetached binary (SH) with two components of almost
equal radii according to \citet{Lazaro2009}, so we have used its A$_1$
value as an upper limit for the depth of SH systems primary minimum.

\section{Application of the classification algorithm}\label{app:application}
Application of the classification method is performed with two main stages. 
First stage is the determination of possible evolutionary classes based on data
from Table~\ref{parameters}. We illustrate this stage for unclassified system DP CMa.
\begin{table*}
\caption{Performing of the classification method for unclassified binary DP~CMa}
\label{determ}
\begin{tabular}{@{}l c c c c c c c l}\hline
\multicolumn{9}{c}{\textbf{DP~CMa}}\\
& A$_1$, mag & A$_2$, mag& $\Delta$A, mag & P, days& Sp$_1$  & Sp$_2$ & $\Delta$Sp$^3$ & MT$^4$\\
&\bf 0.90&\bf 0.30&\bf 0.60&\bf 3.388&\bf K2V&\bf M2V&\bf 1&\bf EA\\[3mm]
Class$^1$& A$_1^{~~2}$ & A$_2^{~~2}$& $\Delta$A$^2$ & P& Sp$_1$  & Sp$_2$ & $\Delta$Sp & MT\\
\multicolumn{9}{l}{Detached binaries}\\[2mm]
DM (190)&  $1.10$ &  & $0.81$ & $[0.4;36]$ & O5--M4.5;IV-VI & O5--M4.5;IV-VI & up to 1.5 & EA, EB, E\\
DR (25)  & $1.56$ & $0.35$ & & $[1.9;26]$ &\cellcolor[gray]{0.5}A8--G6;IV-V & &  &\\
DGE (8)& $0.65$ & $0.34$ & & $[1.6;35]$ &\cellcolor[gray]{0.5}WR3--B2;I,III & && \\
DGL (16)&  $2.32$ &\cellcolor[gray]{0.5}$0.20$ & &  & & & &\\
DW (14)& $6.00$ &\cellcolor[gray]{0.5}$0.20$ & &  &  & &  &\\
D2S (5)&  $6.22$ &  --$^{~~5}$ & &\cellcolor[gray]{0.5}$[603;6310]$ && & &\\[2mm]
\multicolumn{9}{l}{Semidetached binaries}\\[2mm]
SA (376)&  $3.70$ & $0.60$ & & $[2.1;45]$ &\cellcolor[gray]{0.5}B4--G0;I-V &  &  & \\
SC (5)& $1.36$ & $0.55$ & & $[2.9;22]$ & G8--K4;III-V &\cellcolor[gray]{0.5}K1--K5;III-V & &\\
SH (34)&  $1.65$ & $0.57$ & & $[1.1;16]$ &\cellcolor[gray]{0.5}O8--B4;I,III-V & &  & \\
S2C (33)&  $6.00$ &\cellcolor[gray]{0.5}$0.20$ & &  & &  &  & \\[2mm]
\multicolumn{9}{l}{Contact binaries}\\[2mm]
CB (103)&  $1.22$ & $0.81$ & &\cellcolor[gray]{0.5}$[0.2;1.5]$ &  &  &  & \\
CBF (11)&  $1.00$ & $0.30$ & &\cellcolor[gray]{0.5}$[0.5;0.8]$ &&  &  & \\
CBV (13)&$0.91$ & $0.38$ & &\cellcolor[gray]{0.5}$[0.39;1.0]$ &  & &  & \\
CE (19) &  $0.97$ &  &\cellcolor[gray]{0.5}$0.28$ & &&  & &\\
CWA (115)&\cellcolor[gray]{0.5}$0.81$ &  &&  &  & &  & \\
CWW (123)&  $1.00$ &  &\cellcolor[gray]{0.5}$0.22$ &  &  &  & &\\
CG (4)&  $0.69$ &  &\cellcolor[gray]{0.5}$0.12$ &  &  &  &  &\\ 
\hline
\end{tabular}\par
\medskip
$^1$the evolutionary status and the number of such systems in CEV;
$^2$maximum value;
$^3$the components spectral type difference $\Delta$Sp=Sp$_1$-Sp$_2$
is given in units of a spectral class;
$^4$morphological type of the light curve;
$^5$data on secondary minimum are given in CEV for only one D2S system.
\end{table*}

In the third row of Table~\ref{determ} we show values of its parameters: A$_1$=0.90 mag, A$_2$=0.30 mag, 
$\Delta$A=0.60 mag, orbital period P=3.388 days, spectral type and luminosity class for 
primary (more hotter) component Sp$_1$=K2V,  spectral type and luminosity class 
for secondary component Sp$_2$=M2V, the components spectral type difference 
(which is given in units of a spectral class) $\Delta$Sp=Sp$_1-$Sp$_2$=1, and 
morphological type of the light curve MT='EA'.

(or, strictly speaking, we may not) assign to DP CMa, basing on its A$_1$ value.
In the second column of Table~\ref{determ} maximum values of A$_1$ for each
of our classes are listed. Depth of the primary minimum of DP CMa (0.9 mag) is larger
than maximum possible value of A$_1$ of CWA class (0.81 mag). We mark that
cell with the gray colour, and we remove CWA class from further
consideration (other cells of 'CWA' row are empty).

In the second and third steps we determine what classes can not be assigned,
comparing DP CMa's A$_2$ and $\Delta$A values with corresponding maximum values
of remaining evolutionary classes. the reason why we use one of 
these values for different classes is explained in the text (see Sect. 2).

  The second step is to determine what classes are unsuitable for DP CMa 
  basing on its A$_2$ value. We compare A$_2$=0.30 mag of DP CMa with maximum 
  value of A$_2$ of DR, DW, DG*, S*, CB* classes which are listed in third
  column of Table~\ref{determ}. We find that we can exclude from further
  analysis DGL, DW, S2C classes because of A$_2$ value of DP CMa is larger 
  than maximum possible values of A$_2$ of these classes. We mark cells 
  with these A$_2$ with the grey colour again and delete DW, DGL and 
  S2C classes from consideration (other cells in rows 'DW', 'DGL' and 'S2C' 
  are empty).

  In the third step we compare $\Delta$A of our binary with maximum possible 
  value of $\Delta$A of DM, CE, CWW and CG classes. As can be seen from Table~\ref{determ}
  CE, CWW and CG classes should be removed from next steps. We mark
  corresponding cells with the gray colour.

  In the fourth step we determine the possible classes based on period 
  value. We compare orbital period of DP CMa with intervals of possible
  periods for each of the remaining classes. It can be clearly seen 
  that P=3.388 days is longer than the upper limit of interval of 
  periods of CB* classes. We mark these unsuitable periods with grey 
  colour and delete CB* classes from our analysis. D2S class is also 
  impossible for DP CMa because period of this binary is much shorter 
  than lower limit of interval of periods of D2S class.
  
  After these four steps we see that the following evolutionary classes
  can be assigned to DP CMa system: DM, DR, DGE, SA, SC or SH. 

  In the fifth step we check what classes can be assigned, basing on value
  of spectral type of the more hotter component. It can be seen from Table~\ref{determ}
  that only two classes remain, namely DM and SC. Interval of values of
  Sp$_1$ for DR, DGE, SA and SH classes are all unsuitable for DP CMa. We 
  mark the unsuitable intervals with the gray and delete these classes
  from the next steps.

  In the sixth step we compare spectral type of the secondary of DP CMa 
  with interval of possible spectral types of DM and SC classes. 
  Only DM class remains.

  Then we check $\Delta$Sp (step 7) of DP CMa with one that is possible for DM 
  class and also compare values of morphological type in the step 8.

  At the end of this procedure we can classify DP CMa as a DM binary. We
  derive only one possible class so our classification is unique. 
  There is no necessity in MP calculation.

  However, if, as a result of the first stage, more than one class
  can be assigned to a system, we must estimate the MP value
  for each of the possible classes.
  For example let us imagine that there is no information in literature
  about spectral class of the secondary of DP UMa. 
  In this case our classification procedure (first stage, see above) 
  misses steps 6 and 7. As can be seen from Table~\ref{determ} we would have two
  possibilities: DM and SC classes. To choose one of them we
  should execute the second stage and calculate MP value.

  In the first stage we consider binary as a point in the N-dimensional
  space (here N is the number of parameters used for classification 
  in the first stage) and compare its location with location of areas,
  populated with systems of known evolutionary classes. We don't take 
  into account any of the possible observational errors for each of the 
  parameter that we use for classification of the binary.

  The second stage is the estimation of MP value for those binaries which 
  were classified ambiguously, i.e., when we derived more than one class in the 
  first stage. 

\section{List of cluster binaries}\label{app:cluster}
\begin{table*}
\setlength{\tabcolsep}{3pt}
\noindent
\caption{Previously unstudied cluster binaries}
\label{clustermembers}
\begin{tabular}{l l l p{3cm}|l l l p{3cm}}\hline
GCVS      & Cluster & Predicted& Notes & GCVS & Cluster & Predicted& Notes\\
name      & 		&  class   &       & name &         & class    &      \\
V426 Aur  & NGC 1907      & DM  & &
CN Cru    & NGC 4755      & DM  & \\
EV Cnc    & NGC 2682      & CWA & probably CB \citep{Yakut2009} &
DP Cru    & NGC 4609      & DM  & \\
HS Cnc    & NGC 2682      & CWA & &
V2031 Cyg & NGC 6913      & SA  & \\
RV CVn    & NGC 5272      & CB  & &
V2108 Cyg & Roslund 5     & SA  & \\
FF CMa    & Collinder 132 & DM  & &
V2388 Cyg & NGC 6819      & CWW & field star?\\
MS CMa    & Collinder 132 & DM  & &
TZ Lac    & NGC 7243      & SA  & \\
MX CMa    & NGC 2362      & DM  & &
V684 Mon  & NGC 2264      & DM  & \\
QU CMa    & NGC 2354      & CWA & blue straggler? &
V396 Nor  & NGC 6025      & DM  & \\
V422 CMa  & NGC  2362     & DM  & &
V405 Nor  & Loden 2158    & SA  & \\
tau CMa   & NGC 2362      & DGE & multiple, see short description in \citet{Zasche2009}&
AY Per    & Melotte 20    & SA  & \\
GV Car    & NGC 3532      & DM  & &
V578 Per  & Melotte 20    & DM  & \\
QZ Car    & Collinder 228 & DGE & rare object &
BP Per    & Melotte 20    & DM  & \\
V356 Car  & NGC 2516      & DM  & &
V572 Per  & Melotte 20    & DM  & \\
V661 Car  & Trumpler 16   & DGE & &
V620 Per  & NGC 884       & DM  & \\
V546 Cas  & NGC 103       & DM  & &
V621 Per  & NGC 884       & DGE & detached MS+giant binary according to \citet{Southworth2004} \\
V765 Cas  & NGC 457       & DM  & &
V732 Per  & Melotte 20    & DM  & \\
V969 Cas  & NGC 654       & DM  & &
V888 Per  & Melotte 20    & DM  & \\
V1123 Cas & NGC 581       & DM  & &
V607 Pup  & NGC 2422      & DM  & \\
V1130 Cas & NGC 581       & DM  & &
V792 Sgr  & NGC 6514      & DM  & \\
V1133 Cas & NGC 581       & DM  & & 
V5563 Sgr & NGC 6530      & CE  & \\
AI Cep    & Trumpler 37   & DM  & &
V861 Sco  & Trumpler 24   & DGE & studied but not classified \\
IO Cep    & Trumpler 37   & SA  & &
V1069 Sco & NGC 6242      & DGL & \\
SU Cep    & Trumpler 37   & CB  & \citet{Lu1992} confirm our classification &
V1290 Sco & NGC 6231      & DM  & \\
V427 Cep  & Trumpler 37   & DM  & &
V1292 Sco & NGC 6231      & DGE & classified as detached by \citet{Sana2006}\\
V467 Cep  & NGC 6939      & DM  & &
V1293 Sco & Trumpler 24   & DM  & \\
V470 Cep  & NGC 6939      & DM  & &
V1295 Sco & Trumpler 24   & DM  & \\
V735 Cep  & Trumpler 37   & SA  & field star? &
V1297 Sco & NGC 6231      & DM  & \\
V738 Cep  & Trumpler 37   & DM  & &
MY Ser    & NGC 6604      & DM  & studied rare object \\
V747 Cep  & NGC 7822      & DM  & &
QR Ser    & NGC 6611      & DGE & \\
V767 Cep  & NGC 188       & DM  & &
V343 Vel  & NGC 3228      & SA  & \\
MZ Com    & Melotte 111   & DM  & &
V451 Vel  & Pismis 4      & DM  &\\
\hline
\end{tabular}
\end{table*}
\end{document}